# Reasons to Doubt the Impact of AI Risk Evaluations

**Gabriel Mukobi**
UC Berkeley
gmukobi@berkeley.edu

## Abstract

AI safety practitioners invest considerable resources in AI system evaluations, but these investments may be wasted if evaluations fail to realize their impact. This paper questions the core value proposition of evaluations: that they significantly improve our understanding of AI risks and, consequently, our ability to mitigate those risks. Evaluations may fail to improve understanding in six ways, such as risks manifesting beyond the AI system or insignificant returns from evaluations compared to real-world observations. Improved understanding may also not lead to better risk mitigation in four ways, including challenges in upholding and enforcing commitments. Evaluations could even be harmful, for example, by triggering the weaponization of dual-use capabilities or invoking high opportunity costs for AI safety. This paper concludes with considerations for improving evaluation practices and 12 recommendations for AI labs, external evaluators, regulators, and academic researchers to encourage a more strategic and impactful approach to AI risk assessment and mitigation.

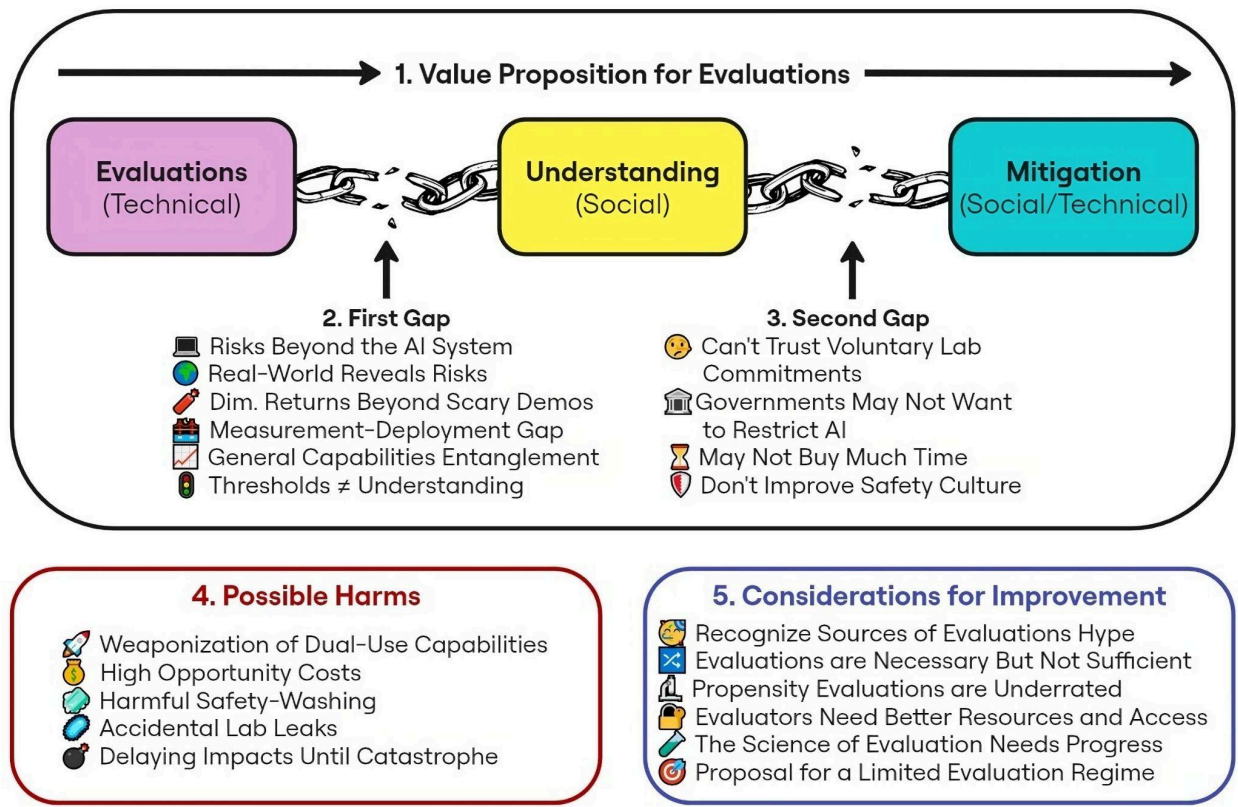

**Figure 1:** The structure of this paper. Evaluations may fail to improve AI risk Understanding or Mitigation and could even cause harm, but evaluations could still be valuable with a few considerations.

# Contents

# 1. The Core Value Proposition of Evaluations

AI safety practitioners currently invest significant talent and resources in AI evaluations, meaning technical methods to test and assess the capabilities or propensities of advanced AI systems.[1] For example, the U.S.[2] and UK[3] AI Safety Institutes include evaluations as one of their top priorities, a significant portion of AI safety technical staff within industry labs are allocated to evaluations, and Apollo Research[4] and METR,[5] which are among the largest AI safety nonprofits, both focus on evaluations.

The core value proposition for AI system evaluations (Figure 1) is that **Evaluations** improve our **Understanding** of the risks of advanced AI systems, and in turn, that improved **Understanding** enables us to better **Mitigate** those risks.[6]

However, I fear that this value proposition may be driven by a techno-solutionist hope that fails to adequately account for a social model of impact.[7] This paper analyzes this possible failure through the links between these three steps: whether **Evaluations** improve **Understanding** (Section 2) and whether **Understanding** improves **Mitigation** (Section 3). Additionally, I discuss possible harms (Section 4), considerations for improving evaluations (Section 5), and recommendations for evaluators (Section 6).

# 2. Evaluations May Fail to Improve Understanding

I critique the first link by describing **six ways** in which AI system **Evaluations** may fail to significantly improve our **Understanding** of a system's risks:

## 2.1 Risks Manifest Beyond the AI System

AI evaluations implicitly focus on risks located within an AI system, such as the system knowing and not refusing requests for instructions on how to build a bioweapon. However, many AI risks manifest through an AI system's interactions with complex systems in the real world, making it especially hard to assess sociotechnical,[8] systemic,[9] or unknown[10] risks with evaluations.

## 2.2 The Real World Reveals Risks

Much of our understanding of AI risks comes from learning about AI's impacts in the real world. Evaluations may not significantly beat the simple baseline of incident reporting from real-world model deployments.[11] We can also learn a lot about risks by collecting information from AI labs, such as through reporting requirements,[12] or embedded government-vetted auditors, and collecting subtle, early-warning signs that external overseers can piece together across the industry.

---

[1] By "Evaluation," I mean the evaluation of AI software systems, not audits of lab practices or other artifacts, which are sometimes also called "evaluations." This paper does discuss related topics such as red teaming or scary demonstrations.
[2] The United States Artificial Intelligence Safety Institute: Vision, Mission, and Strategic Goals | NIST
[3] AI Safety Institute approach to evaluations - GOV.UK
[4] Announcing Apollo Research — Apollo Research
[5] METR
[6] Theories of Change for AI Auditing — Apollo Research, Clarifying METR's Auditing Role, [2305.15324] Model evaluation for extreme risks, AI Safety Institute approach to evaluations - GOV.UK
[7] Safety isn't safety without a social model (or: dispelling the myth of per se technical safety) — AI Alignment Forum
[8] [2310.11986] Sociotechnical Safety Evaluation of Generative AI Systems
[9] [2401.07836] Two Types of AI Existential Risk: Decisive and Accumulative
[10] Emergent Abilities in Large Language Models: An Explainer | Center for Security and Emerging Technology
[11] Preventing Repeated Real World AI Failures by Cataloging Incidents: The AI Incident Database | Proceedings of the AAAI Conference on Artificial Intelligence

## 2.3 Diminishing Warning Returns Beyond Scary Demos

To inform decision-makers about forthcoming AI risks, rigorous and complicated evaluations do not add much more than another baseline of building "scary" demonstrations of specific threat models, but they cost much more. These small demonstrations seemed useful for convincing policymakers at the UK AI Safety Summit of the importance of AI safety,[13] though demonstrators should take care not to misrepresent risks.

## 2.4 Measurement-Deployment Gap

For the foreseeable future, we will likely have a large gap between what evaluations can measure and an AI system's true risks once deployed due to elicitation challenges.[14] Rapid changes to AI systems, such as if future systems undergo continual learning or are connected to the evolving Internet, widen this gap.

## 2.5 General Capabilities Entanglement

Most dangerous capabilities are strongly correlated with general capabilities, so measuring general capabilities—as the ML research community is already incentivized to do—tells most of the story.

Further, Superintelligent AI risks mainly stem from raw general intelligence, not niche domain-specific capabilities, so we should be more concerned with understanding intelligence as time goes on.

## 2.6 Thresholds Are Not Understanding

The current evaluation paradigm is heavily situated within risk management frameworks such as Responsible Scaling Policies (RSPs)[15] that only seek to detect when AI capabilities pass arbitrary capability thresholds.[16] This is chiefly different from a mechanistic understanding of an AI system's risks.

# 3. Understanding May Fail to Improve Mitigation

Next, I discuss **four critiques** of the second link in the value proposition between increased **Understanding** of an AI system's risks as created by evaluations and better **Mitigation** of those risks:

## 3.1 Cannot Trust Voluntary Lab Commitments

Though AI labs are making voluntary commitments now,[17] there is no good reason to expect them to uphold those commitments when they significantly conflict with corporate interests.[18] Especially as AI gets more powerful and race dynamics[19] strengthen, the leading lab will face increased pressure to renege on its promises if it can significantly benefit from deploying its next AI system.

---

[12] [2404.02675] Responsible Reporting for Frontier AI Development
[13] The UK AI Safety Summit - our recommendations — Apollo Research
[14] [2312.07413] AI capabilities can be significantly improved without expensive retraining, Guidelines for capability elicitation | METR's Autonomy Evaluation Resources
[15] Anthropic's Responsible Scaling Policy \ Anthropic, Frontier Safety Framework - Google DeepMind, Preparedness Framework | OpenAI, Responsible Scaling Policies (RSPs) - METR, RSPs are pauses done right — AI Alignment Forum
[16] [2406.14713] Risk thresholds for frontier AI
[17] AI companies make fresh safety promise at Seoul summit, nations agree to align work on risks | AP News
[18] RSPs are pauses done right (Comments) — AI Alignment Forum
[19] [2306.12001] An Overview of Catastrophic AI Risks

## 3.2 Governments May Not Want to Restrict AI

Governments may not be willing to restrict AI systems that evaluations indicate are dangerous due to financial incentives, such as if advanced AI systems have been creating significant economic value.

Additionally, some governments may have a general aversion to slowing innovation, particularly if political powers in charge have established a pro-innovation policy stance.

This lack of political will to act on evaluations may especially reveal itself if pre-deployment evaluations indicate an uncertain possibility of AI risk, but the real world has yet to realize that AI risk (2.2).

## 3.3 Evaluations May Not Buy Much Time

If evaluations indicate an AI system is dangerous, and decision-makers decide to restrict its deployment, it is only a matter of time before AI labs can patch the particular discovered issues and undo that decision. Further, it is not clear if governments can adequately monitor—let alone regulate—internal deployments, such as an AI lab using a frontier system to automate its AI R&D.

It is even more challenging to restrict AI development, which involves many diffuse factors like planning new data centers or testing research ideas, so restricting one dangerous model likely does not significantly affect the timing of subsequent generations of more dangerous models.[20]

## 3.4 Evaluations Alone Do Not Improve Safety Culture

Some hope that requiring evaluations in AI labs will improve the safety culture[21] of those labs.[22] However, this hope seems unfounded, as organizational-wide shifts are required to change safety culture,[23] evaluation teams tend to be siloed within labs, and evaluation requirements could even backfire by creating resentment for safety practices.[24]

# 4. Harm from Evaluations

Beyond these reasons that **Evaluations** may be ineffective at improving **Understanding** and **Mitigation**, Evaluations could be **harmful and increase AI risks** in at least **five cases:**

## 4.1 Weaponization of Dual-Use Capabilities

Dual-use capabilities such as cyber-offense, persuasion, and automated AI R&D[25] are not purely risky—they are also highly desirable for certain actors like national security organizations and AI labs. Evaluations may act as progress measures and triggers for those actors to co-opt dangerous AI systems for their own means.

Even if controlling actors, such as the AI lab that develops a system and the government of the nation that lab is located in, are responsible enough to not weaponize an AI system, evaluations of these dual-use

---

[20] The exception might be second-order resource effects where the revenue and hype from one system deployment helps to fund the next AI system's development.
[21] Building a Culture of Safety for AI: Perspectives and Challenges by David Manheim :: SSRN, Complex Systems for AI Safety [Pragmatic AI Safety #3] — AI Alignment Forum
[22] This may have happened with Anthropic, but that may be due to Anthropic's strong preexisting focus on safety.
[23] Strategy for Culture Change (COS)
[24] When safety culture backfires: Unintended consequences of half-shared governance in a high tech workplace: The Social Science Journal: Vol 46 , No 4
[25] Exclusive: OpenAI working on new reasoning technology under code name 'Strawberry' | Reuters, Examples of AI Improving AI (CAIS)

capabilities may still alert non-responsible actors to the value of stealing that AI system's weights and other artifacts or developing their own similarly powerful AI system.

## 4.2 High Opportunity Costs

Significant AI safety talent and resources are invested in AI evaluations, as described in Section 1. Furthermore, the current evaluation landscape is highly redundant, with many organizations building tests for the same kinds of risks on their own infrastructure. Focusing too much on evaluations may be too costly of a distraction for the AI safety field, especially if evaluations do not buy us a lot more time (3.3).[26] All these people and resources could instead be applied to advancing ML safety and AI governance, making progress on problems that actually mitigate AI risks.

## 4.3 Harmful Safety-Washing

Evaluations might contribute to safety-washing,[27] where non-expert decision-makers are misled into believing that an AI system is safe. This could create a false sense of security,[28] leading to harmful models being deployed.[29]

Relatedly, evaluations could also derisk AI investments,[30] leading to increased investments and AI capabilities acceleration.[31]

## 4.4 Accidental Lab Leaks

Gain-of-function-like capabilities elicitation or intentionally inducing misalignment[32] for scary demonstrations could create unnecessary dangers. As AI systems become more capable, this work could increase the the harm if dangerous models are exfiltrate into the world, either by escaping control on their own or with the aid of human insiders.[33]

This risk, analogous to lab leaks from biosecurity labs intending to study more dangerous pathogens, is made especially significant if AI labs and external evaluators continue to have underdeveloped security practices.[34]

## 4.5 Delaying Impacts Until Catastrophe

Relatedly, if evaluations do not catch all AI risks but rather find more less pernicious or severe risks, then they counterintuitively might increase the harm of the first AI catastrophe. That is, instead of the first point of significant AI harm being a minor incident that would have garnered societal response, incomplete evaluations may catch and prevent those minor incidents but push out the first realized harm to the point of a larger catastrophe that society may be less prepared for.[35]

---

[26] An exception is if time later is much more valuable than time now such that, for example, it is worth it to spend 3 years of AI safety community effort in 2024 to buy 1 year of time in 2028. However, I am skeptical that the time values are so asymmetric and the time we gain later is so long that these kinds of deals may be worth it.
[27] Safetywashing — AI Alignment Forum
[28] When Safety Proves Dangerous
[29] RSPs are pauses done right (Comments) — AI Alignment Forum
[30] Theories of Change for AI Auditing
[31] This may or may not be harmful, partially depending on your view of capability overhangs.
[32] Model Organisms of Misalignment: The Case for a New Pillar of Alignment Research — AI Alignment Forum
[33] Improving the safety of AI evals — LessWrong
[34] Securing AI Model Weights: Preventing Theft and Misuse of Frontier Models | RAND
[35] [2405.19832] AI Safety: A Climb To Armageddon?

This risk depends, however, on both the assumption that AI risk evaluations will fail to detect some of the most severe AI risks and the assumption that societal defense benefits from incremental exposure and adaptation to AI societal impacts.

# 5. Considerations for Improvement

That said, I do not think all evaluations are bad or that some evaluations-like work has no place in the AI safety portfolio. I discuss **six additional considerations** for making the most of evaluations:

## 5.1 Recognize Possible Sources of Evaluation Hype

It is first important to recognize how evaluations became so popular: Evaluations are easy to make continuous progress on, provide ostensibly precise numbers to non-technical decision makers,[36] parallel risk assessment techniques from other disciplines,[37] and are useful for ML capabilities development.[38] However, these are all separate from any reasons that evaluations would be useful for AI risk mitigation.

I especially worry about evaluation hype coming at the expense of AI safety progress. For example, by focusing on evaluations, the UK AI Safety Institute made itself more credible and appealing to the rest of the UK government[39] and may have protected its survival.[40] However, the appearance of value does not necessarily mean this work significantly reduced AI risk better than alternative focuses.

## 5.2 Evaluations are Necessary But Not Sufficient

Evaluations can reveal the presence of risks but not their absence.[41] Some evaluations could be useful for evaluating broad capability levels to trigger tiered security requirements, but we might not expect or gain from much more than the current effort labs are putting into evaluations.

Instead, we may need to flip the burden of proof from evaluating whether assumed-safe AI systems are dangerous into making a case for whether assumed-dangerous AI systems are safe. These safety cases[42] then demand considerable effort into other AI safety interventions that actually reduce risk, such as propensity evaluations, mitigation evaluations,[43] novel mitigation methods, and assurance validation techniques.

## 5.3 Propensity Evaluations are Underrated

Currently, most evaluations focus on capabilities, but we may want to invest more in evaluating the propensities, dispositions, or alignment of AI systems.[44] Propensity evaluations add a crucial component of likelihood to risk assessments, incentivize some progress in alignment, and are harder to weaponize.

---

[36] Richard Ngo on X: "I'm worried that a lot of work on AI safety evals is primarily motivated by "Something must be done. This is something. Therefore this must be done." Or, to put it another way: I judge eval ideas on 4 criteria, and I often see proposals which fail all 4. The criteria:"
[37] Risk assessment - Wikipedia
[38] Let's talk about LLM evaluation
[39] Rishi Sunak on X: "AI is the defining technology of our time and we have a clear strategy to develop it in a safe way that will benefit everyone in the UK. Here's what that looks like 👇
[40] What We Know About the New U.K. Government's Approach to AI | TIME
[41] [2309.01933] Provably safe systems: the only path to controllable AGI
[42] [2403.10462] Safety Cases: How to Justify the Safety of Advanced AI Systems, Affirmative Safety: An Approach to Risk Management for Advanced AI by Akash Wasil, Joshua Clymer, David Krueger, Emily Dardaman, Simeon Campos, Evan Murphy :: SSRN
[43] Mitigation evaluations test how well a mitigation technique reduces some kinds of AI risk. The WMDP Benchmark is an early example of evaluating unlearning mitigations.
[44] [2305.15324] Model evaluation for extreme risks

Propensity evaluations more directly get at our understanding of an AI system and its risks,[45] and as a result, they may need more interpretability progress to start working.[46]

## 5.4 Evaluators Need Better Resources and Access

Industry race dynamics create limitations, such as OpenAI researchers only having one week to evaluate GPT-4o[47] or AI labs withholding pre-deployment access to final models from external auditors,[48] making rigorous evaluations challenging. This can be addressed, but only with strong enough forces to overcome adversarial corporate incentives.

Ideal access might include pre-deployment, transparent-box[49] access to AI systems with both the final versions of models[50] and helpful-only/non-refusal models that are easier to elicit capabilities from, along with lab-developed elicitation and agent tools.

## 5.5 The Science of Evaluation Needs Progress

Insofar as evaluations are beneficial, we can make them more effective by developing rigorous and reproducible evaluation practices. METR has made this its top priority.[51]

However, this is one of the more obvious areas to divert more AI safety research funding to, so it may not be especially neglected soon.

## 5.6 Proposal for a Limited Evaluation Regime

One simplified—but possibly still valuable—evaluation regime not too far from our current RSP-heavy state could involve a light set of evaluations assessing the high-level general capability tier[52,53] of an AI system to trigger heightened security requirements such as safety cases (5.2) and Security Levels.[54] These evaluations could mostly focus on general capabilities rather than many correlated, narrow risks (2.5).

AI labs would do most of the evaluation work, with external parties only really investing in oversight mechanisms such as sending in red teamers to subjectively check the capability tier or reviewing the lab's evaluation practices and infrastructure. AI labs could be incentivized to do decent capability tier evaluations if the increased security requirements mitigate actual business risks,[55] if oversight and legal penalties are strong enough that labs want their internal assessments to be accurate, or if additional security requirements also come with help from government security organizations for achieving those requirements.

# 6. Recommendations

I end with 12 recommendations for possible actions different AI labs, external evaluators, regulators, and academics could take to reduce the failings of and improve AI evaluations.

---

[45] Towards understanding-based safety evaluations — AI Alignment Forum
[46] A transparency and interpretability tech tree — AI Alignment Forum
[47] OpenAI employees say it 'failed' its first test to make its AI safe - The Washington Post
[48] AI companies aren't really using external evaluators - AI Lab Watch
[49] [2401.14446] Black-Box Access is Insufficient for Rigorous AI Audits
[50] GPT-4 System Card (OpenAI)
[51] Clarifying METR's Auditing Role — AI Alignment Forum
[52] For example, AI Safety Levels (ASL) in Anthropic's RSP.
[53] [2406.14713] Risk thresholds for frontier AI
[54] Securing AI Model Weights: Preventing Theft and Misuse of Frontier Models | RAND
[55] For example, consider the Financial impact of the Boeing 737 MAX groundings.

## 6.1 AI Development Labs

1. **Credibly Commit:** Much of the value of evaluations depends on whether whichever AI lab is in the lead will uphold its commitments (3.1). Labs could establish better internal governance mechanisms that could more credibly incentivize them to honor their commitments.[56]
2. **Provide Resources and Access:** They can also provide external evaluators with appropriate resources and access to AI systems (5.4). Labs could directly provide evaluation APIs to let government and third-party evaluators securely use some of the labs' internal tools, such as agent scaffolding, capability elicitation, and grading tools.[57] Labs can also use their considerable financial and computational resources to support the evaluation ecosystem.[58]
3. **Share Evaluation Infrastructure:** Finally, labs can reduce redundancy and increase transparency by sharing much more of their evaluation infrastructure.[59] For example, they could open-source most of their infrastructure[60] or privately share it with trusted government and third-party evaluators. This also increases the transparency of AI lab evaluation methods, allowing for greater scrutiny and trust. Sharing has low downsides, as internal evaluations are already susceptible to exploitation by model developers within AI labs, unless labs have high levels of siloing between evaluation and development teams.

## 6.2 Government and Third-Party Evaluators

4. **Specialize:** To reduce redundancy, AI Safety Institutes and similar organizations might specialize in evaluations that need special government resources, such as national security risks requiring security clearances.[61] Third-party evaluators might specialize in novel risks that labs are less incentivized to work on.[62]
5. **Cooperate on Standards and Sharing:** External evaluators can also focus on planning international cooperations with each other to form globally consistent AI safety standards that reduce the costs and increase the likelihood of compliance and share their work to reduce redundancy.[63]
6. **Create External Oversight:** These groups could oversee AI lab evaluations and scrutinize lab commitments[64] to incentivize greater accountability. Government-vetted experts could be internal auditors embedded within AI labs to leverage heightened transparency.
7. **Build Scary Demos:** To communicate risks to decision-makers, scary demos may better use resources than expensive rigorous evaluations (2.3).
8. **Advance the Science of AI Safety Beyond Evaluations:** Last, external evaluators could advance the science of AI safety.[65] By this, I mean not just the science of evaluations but also making

---

[56] The Windfall Clause is an early but highly non-binding attempt at similar goals.
[57] [2401.14446] Black-Box Access is Insufficient for Rigorous AI Audits
[58] A new initiative for developing third-party model evaluations \ Anthropic
[59] [2404.14068] Holistic Safety and Responsibility Evaluations of Advanced AI Models
[60] OpenAI and DeepMind have made progress in open-sourcing a limited amount of infrastructure, though I expect they could share even more.
[61] Recommendations for the next stages of the Frontier AI Taskforce — Apollo Research
[62] METR started to do this with autonomous adaptation and replication, though that seems less of a priority for METR and more represented within the labs now. Apollo has been specializing in deception evaluations that may use interpretability tools.
[63] The global network of AI Safety Institutes, the U.S.-UK AISI partnership, and the UK AISI open-sourcing its Inspect evaluations infrastructure are all positive signs of this. However, much work remains to figure out the details of efficient resource sharing partnerships.
[64] Commitments - AI Lab Watch
[65] Strategic Vision | NIST

fundamental scientific progress on other technical, governance, and sociotechnical AI safety problems.[66]

## 6.4 AI Regulators

9. **Require Lab Cooperation:** Government regulators can help enforce all the recommendations for AI labs in Section 6.1, increasing the likelihood that AI labs uphold their commitments, provide appropriate resources and access, and share evaluation infrastructure.
10. **Clarify Protections for Lab Cooperation:** Complementarily, regulators could work with AI labs to clarify legal grey zones and carve out protections for specific instances of lab cooperation. This might include anti-trust laws, protections for sharing information, or protections for giving access to external evaluators.[67]

## 6.5 Academic Researchers

11. **Advance the Science of Evaluation:** Researchers can target propensity evaluations (5.3), predicting properties of future AI systems before they are developed,[68] automated evaluations, how to update dynamic evaluations,[69] and other science of evaluation questions (5.5).[70]
12. **Develop Better Threat Models:** Similarly, many evaluation development challenges are bottlenecked by solid threat modeling.[71] Even one person could make significant progress here, improving the efficiency and effectiveness of evaluations built on better threat models.

# 7. Conclusion

There are many reasons to suspect AI **Evaluations** may fail to improve our **Understanding** of AI risks, that the **Understanding** we get from evaluations may fail to improve our **Mitigation** of those risks, and that evaluations could even be harmful. However, evaluations do not seem entirely doomed—we just need to carefully consider how they should fit into a healthy and diverse AI safety portfolio.

AI risk evaluation is a new and rapidly evolving field, so some of these points may be less accurate in different contexts or over time. Ultimately, I aim for this paper to inspire AI safety practitioners to think about the value proposition of evals more deliberately and decide whether and how AI safety talent and resources might be better spent on other means to reduce AI risk.

# Acknowledgments

Many thanks to Dan Hendrycks, David Krueger, and Patricia Paskov for helpful comments and discussions that informed this paper. Mistakes are my own.

---

[66] [2404.09932] Foundational Challenges in Assuring Alignment and Safety of Large Language Models, [2109.13916] Unsolved Problems in ML Safety, [2407.14981] Open Problems in Technical AI Governance
[67] [2403.04893] A Safe Harbor for AI Evaluation and Red Teaming
[68] [2406.04391] Why Has Predicting Downstream Capabilities of Frontier AI Models with Scale Remained Elusive?, [2405.10938] Observational Scaling Laws and the Predictability of Language Model Performance
[69] [2405.10632] Beyond static AI evaluations: advancing human interaction evaluations for LLM harms and risks
[70] See Section 3.3 of [2404.09932] Foundational Challenges in Assuring Alignment and Safety of Large Language Models for more open problems.
[71] Threat Models - AI Alignment Forum

# References

*A new initiative for developing third-party model evaluations*. (n.d.). Retrieved July 22, 2024, from https://www.anthropic.com/news/a-new-initiative-for-developing-third-party-model-evaluations

*AI companies aren't really using external evaluators*. (n.d.). Retrieved July 22, 2024, from https://ailabwatch.org/blog/external-evaluation/

*AI companies make fresh safety promise at Seoul summit, nations agree to align work on risks*. (2024, May 21). AP News. https://apnews.com/article/south-korea-seoul-ai-summit-uk-2cc2b297872d860edc60545d5a5cf598

*AI Safety Institute approach to evaluations*. (n.d.). GOV.UK. Retrieved July 22, 2024, from https://www.gov.uk/government/publications/ai-safety-institute-approach-to-evaluations/ai-safety-institute-approach-to-evaluations

*AI Safety Institute releases new AI safety evaluations platform*. (n.d.). GOV.UK. Retrieved July 22, 2024, from https://www.gov.uk/government/news/ai-safety-institute-releases-new-ai-safety-evaluations-platform

*Announcing Apollo Research*. (n.d.). Apollo Research. Retrieved July 22, 2024, from https://www.apolloresearch.ai/blog/announcing-apollo-research

*Anthropic's Responsible Scaling Policy*. (n.d.). Retrieved July 22, 2024, from https://www.anthropic.com/news/anthropics-responsible-scaling-policy

Anwar, U., Saparov, A., Rando, J., Paleka, D., Turpin, M., Hase, P., Lubana, E. S., Jenner, E., Casper, S., Sourbut, O., Edelman, B. L., Zhang, Z., Günther, M., Korinek, A., Hernandez-Orallo, J., Hammond, L., Bigelow, E., Pan, A., Langosco, L., … Krueger, D. (2024). *Foundational Challenges in Assuring Alignment and Safety of Large Language Models* (arXiv:2404.09932). arXiv. https://doi.org/10.48550/arXiv.2404.09932

*Autonomous replication threat models [draft in progress]*. (n.d.). Google Docs. Retrieved July 22, 2024, from https://docs.google.com/document/d/1be2HNkxPoH0P-q8SDsGrq4xMblPxlpYo9md_zEeSzl0/edit?usp=embed_facebook

Barnes, B. (2024). *Clarifying METR's Auditing Role*. https://www.alignmentforum.org/posts/yHFhWmu3DmvXZ5Fsm/clarifying-metr-s-auditing-role

Booth, H. (2024, July 12). *What We Know About the New U.K. Government's Approach to AI*. TIME. https://time.com/6997876/uk-labour-ai-kyle-starmer/

Cappelen, H., Dever, J., & Hawthorne, J. (2024). *AI Safety: A Climb To Armageddon?* (arXiv:2405.19832). arXiv. https://doi.org/10.48550/arXiv.2405.19832

Casper, S., Ezell, C., Siegmann, C., Kolt, N., Curtis, T. L., Bucknall, B., Haupt, A., Wei, K., Scheurer, J., Hobbhahn, M., Sharkey, L., Krishna, S., Von Hagen, M., Alberti, S., Chan, A., Sun, Q., Gerovitch, M., Bau, D., Tegmark, M., … Hadfield-Menell, D. (2024). Black-Box Access is Insufficient for Rigorous AI Audits. *The 2024 ACM Conference on Fairness, Accountability, and Transparency*, 2254–2272. https://doi.org/10.1145/3630106.3659037

*Commitments*. (n.d.). Retrieved July 22, 2024, from https://ailabwatch.org/resources/commitments/

Critch, A. (2024). *Safety isn't safety without a social model (or: Dispelling the myth of per se technical safety)*. https://www.alignmentforum.org/posts/F2voF4pr3BfejJawL/safety-isn-t-safety-without-a-social-model-or-dispelling-the

Davidson, T., Denain, J.-S., Villalobos, P., & Bas, G. (2023). *AI capabilities can be significantly improved without expensive retraining* (arXiv:2312.07413). arXiv. https://doi.org/10.48550/arXiv.2312.07413

Edwards, M., & Jabs, L. B. (2009). When safety culture backfires: Unintended consequences of half-shared governance in a high tech workplace. *The Social Science Journal*, *46*(4), 707–723. https://doi.org/10.1016/j.soscij.2009.05.007

Hubinger, E. (2022). *A transparency and interpretability tech tree*. https://www.alignmentforum.org/posts/nbq2bWLcYmSGup9aF/a-transparency-and-interpretability-tech-tree

Hubinger, E. (2023). *Towards understanding-based safety evaluations*. https://www.alignmentforum.org/posts/uqAdqrvxqGqeBHjTP/towards-understanding-based-safety-evaluations

Hubinger, E. *RSPs are pauses done right—AI Alignment Forum*. Retrieved July 22, 2024, from https://www.alignmentforum.org/posts/mcnWZBnbeDz7KKtjJ/rsps-are-pauses-done-right#comments

Hubinger, E., Schiefer, N., Denison, C., & Perez, E. (2023, August 7). *Model organisms of misalignment: The case for a new pillar of alignment research - ai alignment forum*. AI Alignment Forum. https://www.alignmentforum.org/posts/ChDH335ckdvpxXaXX/model-organisms-of-misalignment-the-case-for-a-new-pillar-of-1

*Examples of AI Improving AI*. (n.d.). Retrieved July 22, 2024, from https://ai-improving-ai.safe.ai/

FHI, F. of H. I.-. (2020, January 30). *The Windfall Clause: Distributing the Benefits of AI*. The Future of Humanity Institute. http://www.fhi.ox.ac.uk/

Financial impact of the Boeing 737 MAX groundings. (2024). In *Wikipedia*. https://en.wikipedia.org/w/index.php?title=Financial_impact_of_the_Boeing_737_MAX_groundings&oldid=1228175994

*Google-deepmind/dangerous-capability-evaluations*. (2024). [Python]. Google DeepMind. https://github.com/google-deepmind/dangerous-capability-evaluations (Original work published 2024)

*Guidelines for capability elicitation*. (n.d.). METR's Autonomy Evaluation Resources. Retrieved July 22, 2024, from https://metr.github.io/autonomy-evals-guide/elicitation-protocol/

H, D., & ThomasW. (2022). *Complex Systems for AI Safety [Pragmatic AI Safety #3]*. https://www.alignmentforum.org/posts/n767Q8HqbrteaPA25/complex-systems-for-ai-safety-pragmatic-ai-safety-3

Hendrycks, D., Carlini, N., Schulman, J., & Steinhardt, J. (2022). *Unsolved Problems in ML Safety* (arXiv:2109.13916). arXiv. https://doi.org/10.48550/arXiv.2109.13916

Hendrycks, D., Mazeika, M., & Woodside, T. (2023). *An Overview of Catastrophic AI Risks* (arXiv:2306.12001). arXiv. https://doi.org/10.48550/arXiv.2306.12001

Ibrahim, L., Huang, S., Ahmad, L., & Anderljung, M. (2024). *Beyond static AI evaluations: Advancing human interaction evaluations for LLM harms and risks* (arXiv:2405.10632). arXiv. https://doi.org/10.48550/arXiv.2405.10632

*Introducing the Frontier Safety Framework*. (2024, May 14). Google DeepMind. https://deepmind.google/discover/blog/introducing-the-frontier-safety-framework/

Jones, A. (2020). *Are we in an AI overhang?* https://www.alignmentforum.org/posts/N6vZEnCn6A95Xn39p/are-we-in-an-ai-overhang

JustinShovelain, & Elliot_Mckernon. (2023). *Improving the safety of AI evals*. https://www.lesswrong.com/posts/XCRsg2ZnHBNAN862T/improving-the-safety-of-ai-evals

Kasirzadeh, A. (2024). *Two Types of AI Existential Risk: Decisive and Accumulative* (arXiv:2401.07836). arXiv. https://doi.org/10.48550/arXiv.2401.07836

Koessler, L., Schuett, J., & Anderljung, M. (2024). *Risk thresholds for frontier AI* (arXiv:2406.14713). arXiv. https://doi.org/10.48550/arXiv.2406.14713

Kolt, N., Anderljung, M., Barnhart, J., Brass, A., Esvelt, K., Hadfield, G. K., Heim, L., Rodriguez, M., Sandbrink, J. B., & Woodside, T. (2024). *Responsible Reporting for Frontier AI Development* (arXiv:2404.02675). arXiv. https://doi.org/10.48550/arXiv.2404.02675

*Let's talk about LLM evaluation*. (n.d.). Retrieved July 22, 2024, from https://huggingface.co/blog/clefourrier/llm-evaluation

Li, N., Pan, A., Gopal, A., Yue, S., Berrios, D., Gatti, A., Li, J. D., Dombrowski, A.-K., Goel, S., Phan, L., Mukobi, G., Helm-Burger, N., Lababidi, R., Justen, L., Liu, A. B., Chen, M., Barrass, I., Zhang, O., Zhu, X., … Hendrycks, D. (2024). *The WMDP Benchmark: Measuring and Reducing Malicious Use With Unlearning* (arXiv:2403.03218). arXiv. https://doi.org/10.48550/arXiv.2403.03218

Manheim, D. (2023). *Building a Culture of Safety for AI: Perspectives and Challenges* (SSRN Scholarly Paper 4491421). https://doi.org/10.2139/ssrn.4491421

McGregor, S. (2021). Preventing Repeated Real World AI Failures by Cataloging Incidents: The AI Incident Database. *Proceedings of the AAAI Conference on Artificial Intelligence*, *35*(17), Article 17. https://doi.org/10.1609/aaai.v35i17.17817

*METR*. (n.d.). Retrieved July 22, 2024, from https://metr.org/

Nevo, S., Lahav, D., Karpur, A., Bar-On, Y., Bradley, H. A., & Alstott, J. (2024). *Securing AI Model Weights: Preventing Theft and Misuse of Frontier Models*. RAND Corporation. https://www.rand.org/pubs/research_reports/RRA2849-1.html

Nosek, B. (n.d.). *Strategy for Culture Change*. Retrieved July 22, 2024, from https://www.cos.io/blog/strategy-for-culture-change

*Open Problems in Technical AI Governance | GovAI*. (n.d.). Retrieved July 22, 2024, from https://www.governance.ai/research-paper/open-problems-in-technical-ai-governance

OpenAI, Achiam, J., Adler, S., Agarwal, S., Ahmad, L., Akkaya, I., Aleman, F. L., Almeida, D., Altenschmidt, J., Altman, S., Anadkat, S., Avila, R., Babuschkin, I., Balaji, S., Balcom, V., Baltescu, P., Bao, H., Bavarian, M., Belgum, J., … Zoph, B. (2024). *GPT-4 Technical Report* (arXiv:2303.08774). arXiv. https://doi.org/10.48550/arXiv.2303.08774

*Openai/evals*. (2024). [Python]. OpenAI. https://github.com/openai/evals (Original work published 2023)

Phuong, M., Aitchison, M., Catt, E., Cogan, S., Kaskasoli, A., Krakovna, V., Lindner, D., Rahtz, M., Assael, Y., Hodkinson, S., Howard, H., Lieberum, T., Kumar, R., Raad, M. A., Webson, A., Ho, L., Lin, S., Farquhar, S., Hutter, M., … Shevlane, T. (2024). *Evaluating Frontier Models for Dangerous Capabilities* (arXiv:2403.13793). arXiv. https://doi.org/10.48550/arXiv.2403.13793

*Preparedness*. (n.d.). Retrieved July 22, 2024, from https://openai.com/preparedness/

*Recommendations for the next stages of the Frontier AI Taskforce*. (n.d.). Apollo Research. Retrieved July 22, 2024, from https://www.apolloresearch.ai/blog/recommendations-for-the-next-stages-of-the-frontier-ai-taskforce

*Reflections on our Responsible Scaling Policy*. (n.d.). Retrieved July 22, 2024, from https://www.anthropic.com/news/reflections-on-our-responsible-scaling-policy

*Responsible Scaling Policies (RSPs)*. (n.d.). Retrieved July 22, 2024, from https://metr.org/blog/2023-09-26-rsp/

Richard Ngo [@RichardMCNgo]. (2024, July 18). *I'm worried that a lot of work on AI safety evals is primarily motivated by "Something must be done. This is something. Therefore this must be done." Or, to put it another way: I judge eval ideas on 4 criteria, and I often see proposals which fail all 4. The criteria:* [Tweet]. Twitter. https://x.com/RichardMCNgo/status/1814049093393723609

Rishi Sunak [@RishiSunak]. (2023, June 12). *AI is the defining technology of our time and we have a clear strategy to develop it in a safe way that will benefit everyone in the UK. Here's what that looks like 👇 https://t.co/jur7TCS84U* [Tweet]. Twitter. https://x.com/RishiSunak/status/1668169727552765954

Risk assessment. (2024). In *Wikipedia*. https://en.wikipedia.org/w/index.php?title=Risk_assessment&oldid=1235318307

Ruan, Y., Maddison, C. J., & Hashimoto, T. (2024). *Observational Scaling Laws and the Predictability of Language Model Performance* (arXiv:2405.10938). arXiv. https://doi.org/10.48550/arXiv.2405.10938

Schaeffer, R., Schoelkopf, H., Miranda, B., Mukobi, G., Madan, V., Ibrahim, A., Bradley, H., Biderman, S., & Koyejo, S. (2024). *Why Has Predicting Downstream Capabilities of Frontier AI Models with Scale Remained Elusive?* (arXiv:2406.04391). arXiv. https://doi.org/10.48550/arXiv.2406.04391

Scholl, A. (2022). *Safetywashing*. https://www.alignmentforum.org/posts/xhD6SHAAE9ghKZ9HS/safetywashing

Shevlane, T., Farquhar, S., Garfinkel, B., Phuong, M., Whittlestone, J., Leung, J., Kokotajlo, D., Marchal, N., Anderljung, M., Kolt, N., Ho, L., Siddarth, D., Avin, S., Hawkins, W., Kim, B., Gabriel, I., Bolina, V., Clark, J., Bengio, Y., … Dafoe, A. (2023). *Model evaluation for extreme risks* (arXiv:2305.15324). arXiv. https://doi.org/10.48550/arXiv.2305.15324

Strategic Vision. (2024). *NIST*. https://www.nist.gov/aisi/strategic-vision

Street, F. (2020, May 25). *When Safety Proves Dangerous*. Farnam Street. https://fs.blog/safety-proves-dangerous/

Tegmark, M., & Omohundro, S. (2023). *Provably safe systems: The only path to controllable AGI* (arXiv:2309.01933). arXiv. https://doi.org/10.48550/arXiv.2309.01933

*The UK AI Safety Summit—Our recommendations*. (n.d.). Apollo Research. Retrieved July 22, 2024, from https://www.apolloresearch.ai/blog/the-uk-ai-safety-summit-our-recommendations

*Theories of Change for AI Auditing*. (n.d.). Apollo Research. Retrieved July 22, 2024, from https://www.apolloresearch.ai/blog/theories-of-change-for-ai-auditing

*Threat Models—AI Alignment Forum*. (2021, April 19). https://www.alignmentforum.org/tag/threat-models

Tong, A., Paul, K., & Tong, A. (2024, July 15). Exclusive: OpenAI working on new reasoning technology under code name 'Strawberry.' *Reuters*. https://www.reuters.com/technology/artificial-intelligence/openai-working-new-reasoning-technology-under-code-name-strawberry-2024-07-12/

*U.S. and UK Announce Partnership on Science of AI Safety*. (2024, April 1). U.S. Department of Commerce. https://www.commerce.gov/news/press-releases/2024/04/us-and-uk-announce-partnership-science-ai-safety

*U.S. Secretary of Commerce Gina Raimondo Releases Strategic Vision on AI Safety, Announces Plan for Global Cooperation Among AI Safety Institutes*. (2024, May 21). U.S. Department of Commerce.

https://www.commerce.gov/news/press-releases/2024/05/us-secretary-commerce-gina-raimondo-releases-strategic-vision-ai-safety

Verma, P., Tiku, N., & Zakrzewski, C. (2024, July 12). OpenAI promised to make its AI safe. Employees say it 'failed' its first test. *Washington Post*. https://www.washingtonpost.com/technology/2024/07/12/openai-ai-safety-regulation-gpt4/

Wasil, A., Clymer, J., Krueger, D., Dardaman, E., Campos, S., & Murphy, E. (2024). *Affirmative Safety: An Approach to Risk Management for Advanced AI* (SSRN Scholarly Paper 4806274). https://doi.org/10.2139/ssrn.4806274

Weidinger, L., Barnhart, J., Brennan, J., Butterfield, C., Young, S., Hawkins, W., Hendricks, L. A., Comanescu, R., Chang, O., Rodriguez, M., Beroshi, J., Bloxwich, D., Proleev, L., Chen, J., Farquhar, S., Ho, L., Gabriel, I., Dafoe, A., & Isaac, W. (2024). *Holistic Safety and Responsibility Evaluations of Advanced AI Models* (arXiv:2404.14068). arXiv. https://doi.org/10.48550/arXiv.2404.14068

Weidinger, L., Rauh, M., Marchal, N., Manzini, A., Hendricks, L. A., Mateos-Garcia, J., Bergman, S., Kay, J., Griffin, C., Bariach, B., Gabriel, I., Rieser, V., & Isaac, W. (2023). *Sociotechnical Safety Evaluation of Generative AI Systems* (arXiv:2310.11986). arXiv. https://doi.org/10.48550/arXiv.2310.11986

Woodside, T. (2024, April 16). Emergent Abilities in Large Language Models: An Explainer. *Center for Security and Emerging Technology*. https://cset.georgetown.edu/article/emergent-abilities-in-large-language-models-an-explainer/